\documentclass[a4paper,11pt]{article}
\pdfoutput=1 

\usepackage{jheppub} 

\usepackage[T1]{fontenc} % if needed
\usepackage[normalem]{ulem}
\usepackage{amssymb}
\usepackage{amsmath}
\usepackage{color}
\usepackage{xspace}
\usepackage{appendix}
\usepackage{float}
\usepackage[utf8]{inputenc}
\def\red[#1]{{\color{red}#1}}
\def\nno{\nonumber}
\definecolor{military}{rgb}{0.0, 0.5, 0.0}

\preprint{\begin{flushright}
CERN-TH-2019-014 \\ LA-UR-19-21475
\\ TUM-HEP-1190/19
\end{flushright}}

\title{Momentum-space threshold  resummation  in $tW$ production at the LHC }

\author[a,b]{Chong Sheng Li,}
\author[c]{ Hai Tao Li,}
\author[d]{ Ding Yu Shao}
\author[e,f]{and Jian Wang }
\affiliation[a]{Department of Physics and State Key Laboratory of Nuclear Physics and Technology, Peking University, Beijing 100871, China}
\affiliation[b]{Center for High Energy Physics, Peking University, Beijing 100871, China}
\affiliation[c]{ Theoretical Division, Los Alamos National Laboratory,Los Alamos, NM, 87545, USA}
\affiliation[d]{CERN, Theoretical Physics Department, CH-1211, Geneva 23, Switzerland}
\affiliation[e]{Physik Department T31,  Technische Universit\"at M\"unchen, James-Franck-Stra\ss e~1,
D--85748 Garching, Germany}
\affiliation[f]{School of Physics, Shandong University, Jinan, Shandong 250100, China}
\emailAdd{csli@pku.edu.cn}
\emailAdd{haitaoli@lanl.gov}
\emailAdd{dingyu.shao@cern.ch}
\emailAdd{j.wang@sdu.edu.cn}

\abstract{
We calculate the soft-gluon corrections for $tW$ production to all orders.
The soft limit is defined 
in the pair invariant mass or one particle inclusive kinematic schemes. We find that at NLO the contribution of the soft-gluon effect dominates in the total cross section or the differential distributions. 
After resumming the soft-gluon effect to all orders using the renormalization group equation, 
we find that the NLO+NNLL results
increase the NLO cross sections by  $12\%\sim 17\%$ depending on the scheme and the collider energy. 
Our results are in agreement with
the measurements at the 8 and 13 TeV LHC. 
We also provide predictions for the total cross section  at the 14 TeV LHC.
}

\begin{document} 
\maketitle

\section{Introduction}

The top quark, the most massive elementary particle discovered so far, is playing an important role in testing the standard model (SM) and searching for new physics beyond the SM. At a hadron collider, such as the large hadron collider (LHC), the top quarks can be produced in pairs via strong interaction or in association with a jet or a $W$ boson via weak interactions. The associated production with a $W$ boson offers a particular window to the weak interactions of the top quark and potentially can lead to a direct measurement of the CKM matrix element $V_{tb}$. Besides, $tW$ production is the second largest single top production channel and thus serves as an essential background in search for new physics. So far, the LHC has accumulated a large number of data, based on which the total and differential cross sections of this channel have been measured directly \cite{Aad:2012xca,Aad:2015eto,Aaboud:2016lpj,Aaboud:2017qyi,Chatrchyan:2012zca,Chatrchyan:2014tua,Sirunyan:2018lcp}.

On the other hand, the precise theoretical predictions provide a valid framework in which important information can be extracted from the experimental data. Since the leading-order (LO) cross section of the $W$ boson associated production is proportional to the strong coupling $\alpha_s$ and $|V_{tb}|^2$, it is crucial to understand the value of  $\alpha_s$ in order to extract $V_{tb}$. The inclusion of higher order QCD corrections can help to estimate the factorization and renormalization scale dependence of the cross sections. The next-to-leading order (NLO) QCD corrections have been calculated in refs.~\cite{Giele:1995kr,Zhu:2001hw,Cao:2008af, Kant:2014oha}, and the results including decays of the top quark and the $W$ boson are also available \cite{Campbell:2005bb}. The parton shower effects in this channel have been studied in refs.~\cite{Frixione:2008yi,Re:2010bp,Jezo:2016ujg}.

When considering the higher order corrections for $tW^-$ production, there is one subtlety to deal with. Due to the same $tW^-b$ final state in both the real correction to $tW^-$ and the $t{\bar t}$ production with the decay ${\bar t}\to W^-b$, one needs to find a way to differentiate the two processes or to define the $tW^-$ process properly beyond tree level. There are several proposals on the market \cite{Tait:1999cf, Belyaev:2000me, White:2009yt,Demartin:2016axk}. Here we point out that so far we discuss the $tW^-$ production in the five flavor scheme, i.e., the LO is $gb\to tW^-$. It is also possible to work in the four flavour scheme in which the LO is $gg\to tW^-\bar{b}$ \cite{Frederix:2013gra,Cascioli:2013wga}.

The next-to-next-to-leading order (NNLO) QCD corrections to $tW$ production are not available for now. Though the NNLO $N$-jettiness soft function, one of the ingredients for an NNLO calculation using $N$-jettiness subtraction method, has been computed in refs.~\cite{Li:2016tvb,Li:2018tsq}, the two-loop virtual correction is still the bottleneck because of its dependence on multiple scales. The soft gluon corrections near the threshold have been calculated up to NNNLO based on next-to-next-to-leading logarithms (NNLL) resummation \cite{Kidonakis:2006bu,Kidonakis:2010ux,Kidonakis:2016sjf}, which are considered as an approximation to the full higher order corrections.  The three-loop soft anomalous dimension for $tW$ production was calculated by ref.~\cite{Kidonakis:2019nqa} which can be used to study the full NNNLO threshold effects.

In this paper, we will present the soft gluon corrections to  $tW$ production using the soft-collinear effective theory \cite{Bauer:2000ew,Bauer:2000yr,Bauer:2001ct,Bauer:2001yt,Beneke:2002ph} (see \cite{Becher:2014oda} for a review) which separates the hard contributions with the large momentum transfer and the soft gluon corrections characterized as low energy contributions. Two different definitions of the soft limit are investigated. One is measured by the threshold variable $1-z=1-M_{tW}^2/\hat{s}\to 0$, while the other is given by $s_4=(p_1+p_2-p_t)^2-M_W^2\to 0$. In principle, these two definitions encode the same soft gluon physics in the threshold limits and they only differ by power suppressed corrections. The threshold contributions up to NNLO are obtained  and the resummation is achieved through solving the RG equations of the hard and soft functions. Our results could be taken as an important theoretical input in future experimental analyses.

The paper is organized as follows. In section \ref{sec:factorization}  we show the basic information about the kinematics in this process and the factorization formula of the cross section in the soft limit. The numerical results and relevant discussions are then presented in section \ref{sec:numerical}. We conclude in section \ref{conclu}. The evolution equation of parton distribution functions (PDFs) in the threshold limits and the analytic result of the soft function are given in the appendices.

\section{Factorization and resummation formalism} \label{sec:factorization}

\begin{figure}[t]
 \centering
 \includegraphics[scale=0.4,keepaspectratio=true]{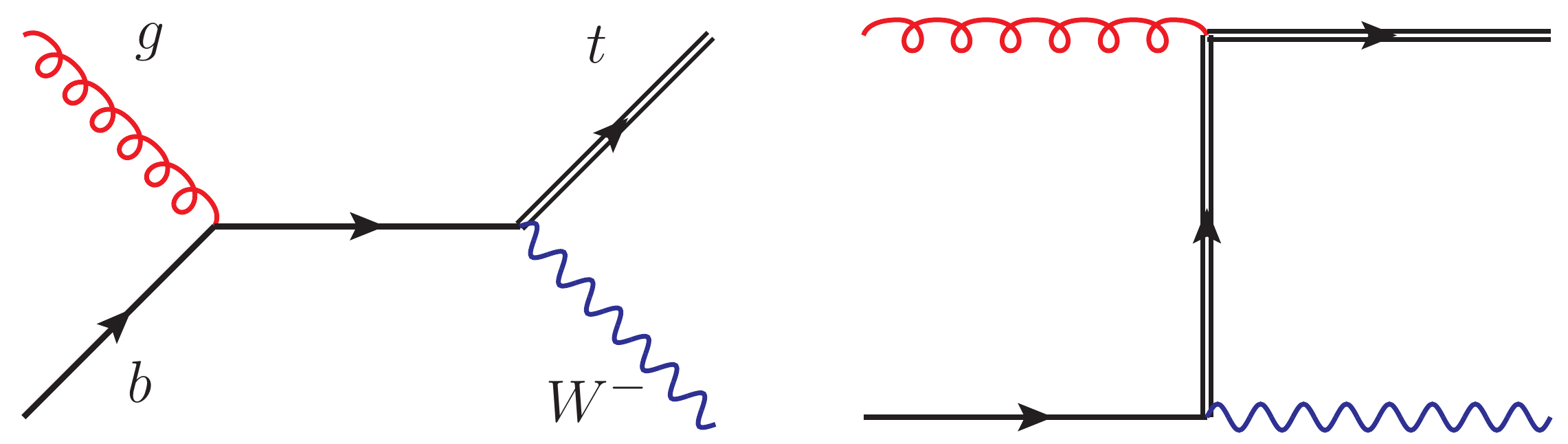}
 \caption{LO Feynman diagrams for $tW^-$ production.}
 \label{fig:lo_dia}
\end{figure}

We consider inclusive stable top quark and $W$ boson associated production at the LHC
\begin{align}
    p(P_1) + p(P_2) \to t(p_3) + W^-(p_4) + X(P_X),
\end{align}
where $X$ denotes all the other possible extra radiations in the final states. In the threshold limit at the leading power we only need to consider the partonic channel  
\begin{align}
    b(p_1) + g(p_2) \to t(p_3) + W^-(p_4) + X(p_X).
\end{align}
The corresponding LO Feynman diagrams are shown in figure~\ref{fig:lo_dia}. 
The partonic kinematic variables are defined to be
\begin{align}\label{kindef}
&\hat{s}=(p_1+p_2)^2,~~\hat{t}_1=(p_1-p_3)^2-m_t^2, ~~ \hat{u}_1=(p_2-p_3)^2-m_t^2, 
\nonumber \\ 
&\hat{t}_1^{\,W}= (p_2-p_4)^2-M_W^2,  ~~~~ \hat{u}_1^W=(p_1-p_4)^2-M_W^2. 
\end{align}
The corresponding variables at hadronic level are
\begin{align}
    s=\hat{s}/x_1/x_2 ,~~ t_1 =  \hat{t}_1 /x_1, ~~ u_1 =  \hat{u}_1 /x_2, ~~ t_1^{\, W} =  \hat{t}_1^{\,W}/x_2, ~~ u_1^W =  \hat{u}_1^W /x_1,
\end{align}
where $x_{1,2}$ are the Bjorken scaling variables. 

In the soft limit or threshold limit, the real emissions are highly constrained, only soft gluons allowed in the final state $X$. This limit would be reached if the invariant mass of the final state $M=\sqrt{(p_3+p_4)^2}$ approaches the initial partonic center-of-mass energy $\sqrt{\hat{s}}$. As a result, the variable  $1-z\equiv 1-M^2/\hat{s}\to 0$ in the threshold limit\footnote{In the central-of-mass frame the energy of the soft radiation is $E_g \approx M(1-z)/2\sqrt{z}$. }, and the perturbative expansion of cross section contains a series of large logarithms $\alpha_s^n\left[\ln^{2n-i} (1-z)/(1-z)\right]_+ ~~(i=1,2,...,2n)$,  which might spoil the convergence of perturbative series. It is our purpose in this work to study the threshold behavior and resum such large logarithms to all orders. Since the soft limit is characterized by the final-state two particles' invariant mass, it is called the pair invariant mass (PIM) scheme. Besides, there is another scheme, called one particle inclusive (1PI) scheme, in which the soft limit is defined by partonic level $s_4\to 0$ with $s_4 \equiv \hat{s}+\hat{t}_1+\hat{u}_1+m_t^2-M_W^2 =(p_4+p_X)^2-M_W^2$. The two schemes measure the soft limit in different ways and the combination of the studies in two schemes provides more complete information on the structure.

In the rest part of this section we briefly show the factorization formula for $tW$ production, which can be derived in a similar way used in the other processes, such as the single top or top quark pair productions \cite{Ahrens:2010zv,Ahrens:2011mw,Zhu:2010mr,Wang:2010ue,Wang:2012dc}. 
In the PIM scheme the cross section in the threshold limit can be factorized to a product of the hard and the soft function
\footnote{This factorization is carried out at the leading power of the threshold variable. At next-to-leading power, the threshold factorization becomes more complicated; see the recent paper \cite{Beneke:2018gvs}.}~\cite{Idilbi:2005ky,Becher:2007ty}, which describes the physics at two different scales, i.e., the large hard scale and the small soft scale, respectively. 
They contain no large logarithms at their intrinsic scales ($\mu_h$ and $\mu_s$ respectively) as expected, since they depend only on a single scale there.
The resummation of all the large logarithms caused by soft gluon effects is achieved by evolving the two functions from the intrinsic scales to a common factorization scale using their renormalization group (RG) equations.  
The RG improved differential cross section can be written as  
\begin{align}\label{eq:pim}
    \frac{d^2\sigma^{\rm PIM}}{dM^2  d\cos\theta } = & \frac{\lambda^{1/2}}{32 \pi s M^2} \sum_{ij} \int_\tau^1 \frac{dz}{z} \int_z^1 \frac{dx}{x} f_{i/p}(x, \mu_f) f_{j/p}(z/x, \mu_f) H(\mu_h) U_{\rm PIM}(\mu_h,\mu_s,\mu_f)  \nonumber \\
    & \hspace{-1.2cm}\times \frac{z^{-\eta}}{(1-z)^{1-2\eta}} \tilde s_{\rm PIM} \left( \ln \frac{M^2(1-z)^2}{z\mu_s^2} + \partial_\eta,\mu_s \right)
     \frac{e^{-2\gamma_{\rm E} \eta}}{\Gamma(2\eta)} \Bigg |_{\eta = (C_A+C_F) a_{\gamma^{\rm cusp}} (\mu_s, \mu_f)},
\end{align}
where polar angle of top quark $\theta$ is defined in the center-of-mass frame, $\lambda = (1-m_t^2/\hat{s}-M_W^2/\hat{s} )^2-4 m_t^2 M_W^2/\hat{s}^2 $, $\tau = M^2/s$ and the other kinematic variables have already been defined in eq.~\eqref{kindef}. 
For convenience we suppress the dependence of hard function $H$, soft function $\tilde s_{\rm PIM}$ and evolution factor $U_{\rm PIM}$ on the kinematic variables. $\tilde{s}_{\rm PIM}$ denotes the soft function defined in the Laplace space. 
The evolution factor embodies RG running from hard and soft scale to factorization scale, which is expressed as
\begin{align}
U_{\rm PIM}(\mu_h, \mu_s, \mu_f) = &\left(\frac{M^2}{\mu_h^2}\right)^{(C_A+ C_F)\, a_{\gamma^{\rm cusp}}(\mu_s,\mu_h)} \exp\Big[ 2(C_A + C_F)S(\mu_h,\mu_s) \nno \\
 & \hspace{2.4cm} + \, a_{\gamma^h}(\mu_s,\mu_h)  + 2\,a_{\gamma^{\phi_q}}(\mu_s,\mu_f) + 2\,a_{\gamma^{\phi_g}}(\mu_s,\mu_f) \Big]. 
\end{align}
where the definitions of function $S$ and $a_\gamma$ and all relevant anomalous dimensions, e.g., $\gamma^{\phi_q}$, can be found in appendix A of ref.~\cite{Ahrens:2010zv}. 
The hard anomalous dimension specific to the $tW$ process is given by~\cite{Ferroglia:2009ep,Ferroglia:2009ii} 
\begin{align}
\gamma^h = 2\left( \gamma^Q + \gamma^q + \gamma^g \right) - \frac{C_A}{2} \gamma^{\rm cusp} \ln \frac{m_t^2 (-\hat s)}{\hat u_1^2} - \left( C_F - \frac{C_A}{2}\right) \gamma^{\rm cusp} \ln \frac{m_t^2 (-\hat s)}{\hat t_1^{\, 2}} . 
\end{align}

In the 1PI scheme,  the soft radiations are characterized via the threshold variable $s_4 \approx  2 p_4 \cdot k $ with $k$ the sum of all the momenta of the soft final state~\cite{Laenen:1998qw,Becher:2009th,Ahrens:2011mw, Wang:2012dc}.  The analysis is simplified by going to the rest frame of the inclusive final state $W+X$ where $|\Vec{p}_W| = \mathcal{O}(s_4/M_W)$. In this frame the energy of the soft radiation is  $E_g \approx s_4/2 M_W$. In the 1PI scheme, the RG improved cross section is 
\begin{align}\label{eq:1pi}
   \frac{d^2\sigma^{\rm 1PI}}{dp_T^2  dy } =& \frac{  1 }{ 16 \pi s} \sum_{ij}\int_{x_1^{\min}}^{1} \frac{dx_1}{x_1} \int_0^{s_4^{\max}}
   \frac{ds_4}{s_4+M_W^2-m_t^2-x_1 t_1}  f_{i/p}(x_1, \mu_f)f_{j/p}(x_2, \mu_f)  \nonumber \\ & 
   \times H(\mu_h)  U_{\rm 1PI}(\mu_h,\mu_s,\mu_f) \tilde{s}_{\rm 1PI}(\partial_\eta,\mu_s) \frac{1}{s_4}\left(\frac{s_4}{M_W \mu_s}\right)^{2\eta} \frac{e^{-2\gamma_{\rm E} \eta}}{\Gamma(2\eta)},
\end{align}
where the integration range is defined as $ x_1^{\min} = (M_W^2-m_t^2-u_1)/(s+t_1)$, $s_4^{\rm max} = m_t^2 - M_W^2 + u_1 + x_1(s+t_1)$, and momentum fraction $x_2$ is defined via $x_1$ and $s_4$ as $x_2 = (s_4+M_W^2-m_t^2-x_1 t_1)/(u_1 + x_1 s)$. The Mandelstam variables are related to the top quark's transverse momentum $p_T$ and rapidity $y$ via $t_1=-\sqrt{s}\,m_{\bot}e^{-y}$, and $u_1=-\sqrt{s}\,m_{\bot}e^{y}$, with $m_{\bot}=\sqrt{\smash[b] p_T^2+m_t^2}$. 
Here the 1PI evolution factor has the form as 
\begin{align}
U_{\rm 1PI}(\mu_h, \mu_s, \mu_f) &=     \left(\frac{M^2}{\mu_h^2}\right)^{(C_A+C_F)a_{\gamma^{\rm cusp}}(\mu_s,\mu_h)}
  \exp\Bigg[ 2(C_A+C_F)S(\mu_h,\mu_s) + \, a_{\gamma^h}(\mu_s,\mu_h) \nno \\
 & \hspace{-2cm}   + 2\,a_{\gamma^{\phi_q}}(\mu_s,\mu_f) + 2\,a_{\gamma^{\phi_g}}(\mu_s,\mu_f)  +a_{\gamma^{\rm cusp}}(\mu_s,\mu_f)\Bigg(C_A \ln\frac{M_W^2\mu_s^2}{\left( \hat{t}_1^{\,W}\right)^2}+C_F \ln\frac{M_W^2\mu_s^2}{\left( \hat{u}_1^W\right)^2}\Bigg) \Bigg]. 
\end{align}
The difference between the RG factors in eq.~(\ref{eq:pim}) and eq.~(\ref{eq:1pi}) arises from the RG equation of the PDFs and soft function. We present the RG evolution of the PDFs in the PIM and 1PI schemes in appendix~\ref{app:pdfs}. The two-loop anomalous dimensions that govern the evolution of the hard and soft functions and thus determine the scale dependent part  are derived from the general structure of the anomalous dimension~\cite{Ferroglia:2009ep, Ferroglia:2009ii}. 
The scale independent part of the hard function has been obtained at NLO using modified MadLoop~\cite{Hirschi:2011pa} which makes uses of Ninja~\cite{Peraro:2014cba}, CutTools~\cite{Ossola:2007ax} and OneLOop~\cite{vanHameren:2009dr} packages. 
We have computed one-loop soft function analytically, which is shown in appendix~\ref{app:soft}. Combining all the ingredients together,  we have checked the RG invariance
\begin{align}
  \frac{d}{d\ln \mu} \left( f_{i/p}\otimes f_{j/p} \otimes H \otimes S_{\rm PIM, 1PI} \right) = 0
    \label{eq:sclindep}
\end{align}
in both of the kinematic schemes.

The NLO and NNLO leading power contributions are obtained by setting the scales  $\mu_h, \mu_s, \mu_f  $ in Eqs.~(\ref{eq:pim}) and (\ref{eq:1pi})  equal. In this way, for PIM scheme we capture all the threshold logarithms $\alpha_s \left[  \ln^n (1-z)/(1-z)\right]_+$ with $n=1,0$  at NLO and $\alpha_s ^2 \left[\ln^n (1-z)/(1-z)\right]_+$ with $n=3,2,1,0$ at NNLO, as well as the scale dependent logarithms predicted by eq.~(\ref{eq:pim}).  
The similar procedure can be applied to obtain the threshold enhanced logarithms for 1PI scheme.  In the following calculations  the approximate NNLO (aNNLO) cross section is defined as 
\begin{align}
    d\sigma({\rm aNNLO } ) = d\sigma({\rm NNLO\  leading } ) + d\sigma({\rm NLO }) - d\sigma({\rm NLO\  leading } ),
    \label{eq:annlo}
\end{align}
where the NLO power suppressed terms in $1-z$ or $s_4$ have been included to give more precise results. We can also match the resummed prediction to the fixed order result by 
\begin{align}
    d\sigma({\rm NLO+NNLL } ) = d\sigma({\rm NNLL } ) + d\sigma({\rm NLO }) - d\sigma({\rm NLO\ leading } )
    \label{eq:nnllnlo}
\end{align}
with the NNLL result given by  eq.~(\ref{eq:pim}) and eq.~(\ref{eq:1pi}).

\section{Numerical results} 
\label{sec:numerical}

To perform the numerical calculation,
the input parameters are set as $m_t=173.3$ GeV, $\Gamma_t=1.5$ GeV, $M_W=80.419$ GeV, $\alpha=1/132.5$ and the Fermi-constant $G_F=1.166390 \times 10^{-5} $ GeV$^{-2}$.
For the LO and NLO calculations we use the CT14 LO and NLO PDF sets~\cite{Dulat:2015mca} as provided by the LHAPDF library~\cite{Buckley:2014ana}, respectively. The aNNLO  and resummed  predictions are obtained using CT14 NNLO PDF sets.   
For fixed-order calculations  the renormalization scale is set to be the same as the factorization scale. 
It is natural to set the default hard scale to be the invariant mass of the top quark and $W$ boson, i.e., $\mu_h=M$, where the hard function contains no large logarithms. 
The soft scale is chosen numerically according to the criterion that the perturbative series of the soft function are well behaved~\cite{Becher:2007ty}.
Explicitly, we find that the ratio of the soft and hard scale is $0.3 \sim 0.4$ in the PIM scheme and
$0.3 \sim 0.5$ in the 1PI scheme at the 8 TeV LHC. 
The default factorization scale has been chosen to be $\mu_f=M$ and $\mu_f=m_t+M_W$ in the PIM and the 1PI schemes, respectively.
The final scale uncertainties are evaluated by varying these scales by a factor of two independently. 

As discussed in the introduction there are several methods to deal with the problem of the interference between the real corrections to $tW$ production and $t\bar{t}$ calculation ~\cite{Tait:1999cf, Belyaev:2000me, Campbell:2005bb,Frixione:2008yi,White:2009yt,Demartin:2016axk} at NLO, such as Diagram Removal (DR), Diagram Subtraction (DS) and b-jet transverse momentum veto. 
The differences between these schemes have been discussed a lot before; see e.g. refs.~\cite{Frixione:2008yi,Re:2010bp}.
Since they are only relevant in the power suppressed channels\footnote{The problem of the interference exists only in the $gg\to tW^-\bar{b}$ or $q\bar{q}\to tW^-\bar{b}$ channel which is at subleading power near the threshold.},
we will not repeat the discussion about their difference in this paper.
We notice that in refs.~\cite{Kidonakis:2010ux,Kidonakis:2016sjf} where the higher order threshold corrections were studied, the power corrections as well as certain leading power logarithmic independent terms are not taken into account. 
In the rest of this section we only show the predictions  of the process $ p p \to tW^-$. The total cross section for $tW^-$ and $\bar{t}W^+$ can be obtained by doubling the results, as demonstrated in ref.~\cite{Frixione:2008yi}. The NLO cross sections in the b-jet veto scheme are evaluated using MCFM~\cite{Campbell:2005bb} with $p_T^{b-{\rm jet}}<50$ GeV. The cross sections in the DS and DR schemes are calculated by POWHEG-BOX~\cite{Alioli:2010xd,Re:2010bp}. 

\begin{table}[t]
    \centering
    \resizebox{\textwidth}{!}{\begin{tabular}{l|c|c|c|c|c|c}
    \hline \hline 
         [pb]   & \multicolumn{3}{c|}{PIM  }   & \multicolumn{3}{c}{1PI}  \\
    \hline 
       &     b-veto     &  DS   &  DR  &   b-veto     &  DS   &  DR   \\
         \hline     
          LO   & \multicolumn{3}{c|}{$6.96_{-6\%}^{+5\%}$   }   & \multicolumn{3}{c}{$7.21^{+5\%}_{-4\%}$}  \\
         \hline     
          NLO $bg$   & \multicolumn{3}{c|}{$12.0_{+2\%}^{-6\%}$   }   & \multicolumn{3}{c}{$11.7_{+4\%}^{-8\%}$}  \\
    \hline 
            NLO leading   & \multicolumn{3}{c|}{$11.2_{+2\%}^{-5\%}$   }   & \multicolumn{3}{c}{$11.5_{+0\%}^{-3\%}$}  \\
    \hline 
     NLO   &  $9.31^{+0\%}_{-1\%}$    & $9.92^{+2\%}_{-2\%}$ &  $10.0^{+2\%}_{-2\%}$ & $9.32^{-1\%}_{+0\%}$    & $10.0^{+1\%}_{-2\%}$   & $10.2^{+2\%}_{-2\%}$   
     \\         
         \hline 
     power corr.  & $-1.87_{-0.25}^{+0.5}$ & $-1.26_{-0.40}^{+0.75}$ & $-1.17_{-0.43}^{+0.76}$  &  $-2.14_{-0.05}^{+0.30}$ &   $-1.46_{-0.21}^{+0.53}$ &  $-1.29_{-0.27}^{+0.63}$
     \\
    \hline \hline
    \end{tabular}}
    \caption{The fixed-order  total cross section and the power corrections  for $tW^-$ production with $\sqrt{s}=8$ TeV. The power corrections are defined as $ d\sigma({\rm NLO }) - d\sigma({\rm NLO\ leading } ) $. The LO results are different in two schemes due to the different choice of the factorization scale. The scale uncertainties are shown.
    }
    \label{tab:nlosigma}
\end{table}

%
%
% figure added 
\begin{figure}
    \centering
    \includegraphics[width=0.6\textwidth]{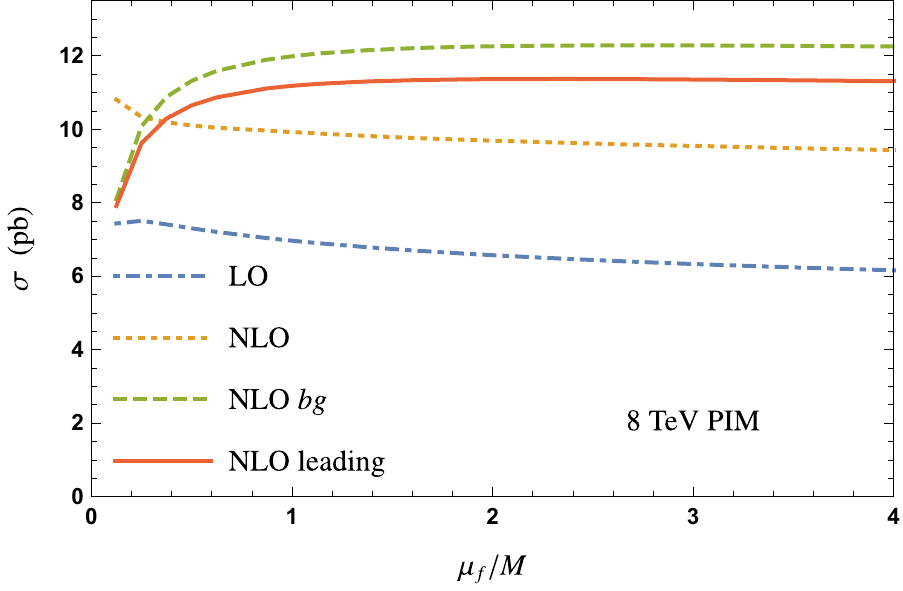}
    \caption{The factorization scale dependence of the cross sections in the PIM scheme for $tW^{-}$ production with $\sqrt{s}=8$ TeV. The NLO result is obtained in the DS scheme. The plots are shown in the region $1/8<\mu_f/M<4$.}
    \label{fig:PIMscale}
\end{figure}

Before presenting the resummed result, we firstly investigate the contribution of the leading power terms.
From table \ref{tab:nlosigma} we can see that the NLO corrections are sizeable,  enhancing the LO result by $29\% \sim 44\%$ depending on 
the different methods to isolate the $tW$ process.
These NLO corrections get contributions from all the $bg$, $gg$ and $qq^\prime$ channels, though at LO only $bg$ channel exists.
Among them,  the $bg$ channel dominates or even surpasses the NLO corrections, as indicated in table \ref{tab:nlosigma} too.
Moreover, the leading power terms of the $bg$ channel can approximate the total  result of the $bg$ channel very well, 
the difference being only $7\%$ and $2\%$  in the PIM and 1PI schemes, respectively.
Since the leading power terms can be obtained from the resummed results as discussed above in eq.(\ref{eq:annlo}),
they can be calculated up to higher orders in $\alpha_s$, namely beyond NLO.
These make up a major part of the full NNLO corrections and can be taken as an approximation of the latter. 
The quality of the approximation could be estimated by looking at the power corrections.
The NNLO results are still unavailable, so we study the NLO ones which are shown in  table \ref{tab:nlosigma}  as well.
It is ready to see that they are negative and around $-20\%\sim -12\%$ in PIM and $-23\%\sim -13\%$ in 1PI kinematic scheme
depending on the methods to deal with the interference problem.
The contributions of the higher order (in $\alpha_s$) power corrections 
can be obtained by calculating the full NNLO QCD corrections or by making use of the next-to-leading power factorization and resummation,
both of which are difficult at the moment and beyond the scope of this paper.

Although the usual way to evaluate the scale uncertainty is to vary the scales by a factor of two, it is also 
interesting to investigate the factorization scale dependence in a larger region. 
From the figure~\ref{fig:PIMscale}, we can see that  the ratio of the NLO over LO result is insensitive to the factorization scale, always in the region $(1.37,1.53)$, when it is varied from $M/8$ to $4M$. This means that there is no clear choice of the factorization scale to ensure fastest convergence.
Moreover, we find that the $bg$ channel is very sensitive to the factorization scale when it is smaller than $M/2$. In order to avoid such a dependence, we have chosen the default factorization scale at $M$.
It can also be seen that the NLO leading power terms dominate the $bg$ channel over a large region.

\begin{figure}[t]
    \centering
    \includegraphics[width=0.49\textwidth]{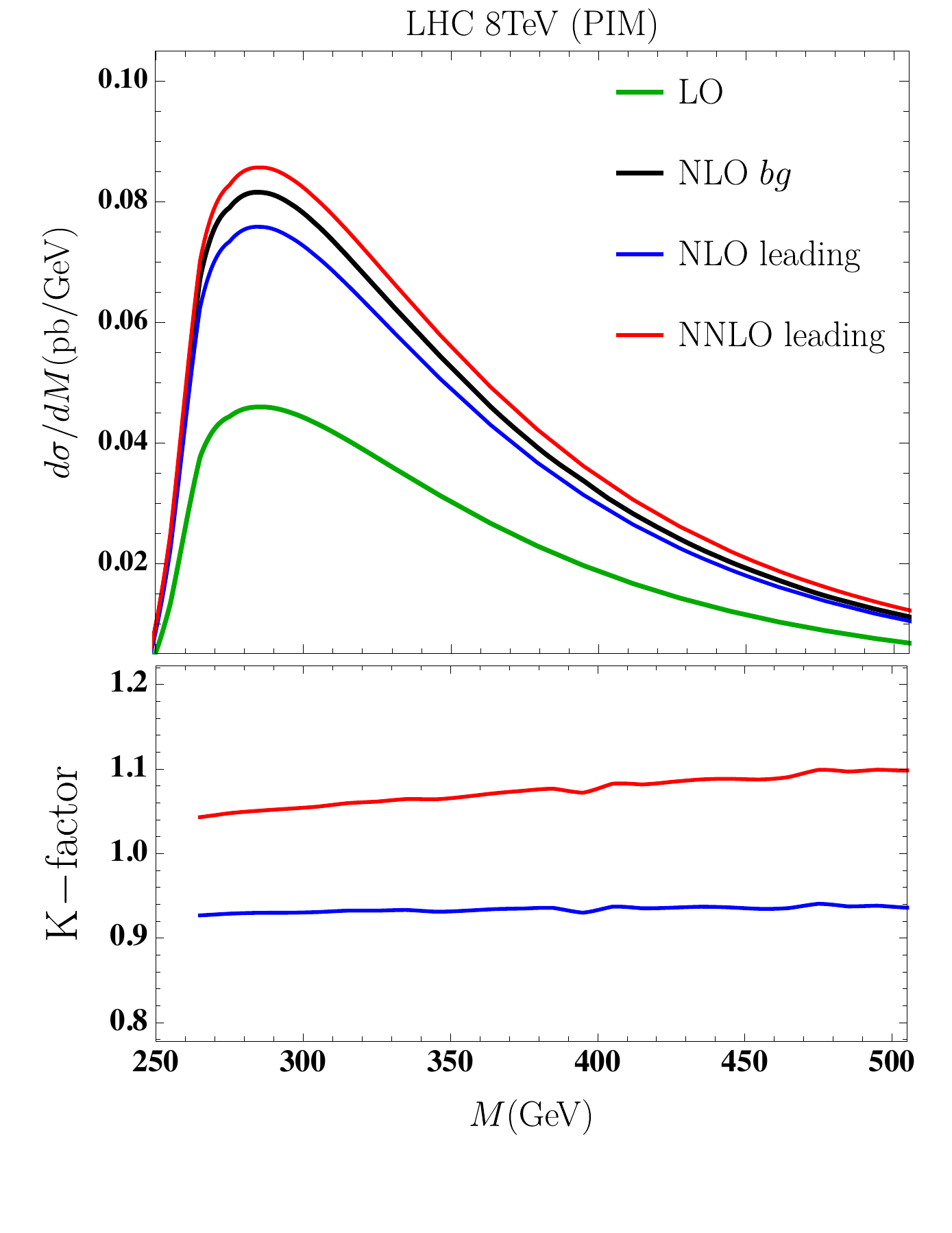}~
    \includegraphics[width=0.49\textwidth]{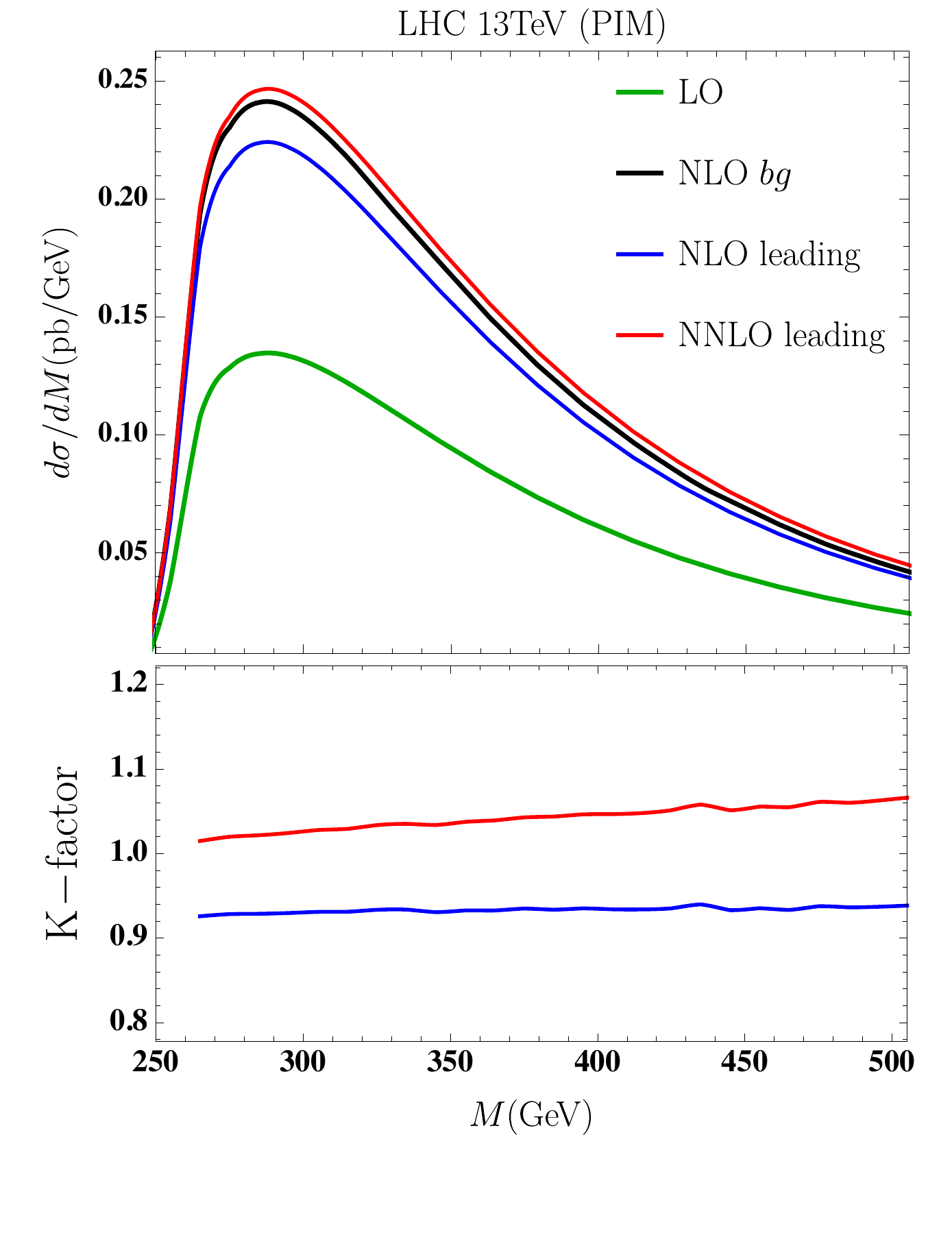}
    \vspace{-1.5cm}
    \caption{Invariant mass distributions in the PIM scheme for $tW^-$ production. 
    In the upper plots, the black lines represent the NLO cross section from $bg$ channel while the blue and red lines are the NLO leading and NNLO leading predictions, respectively. In the bottom plots, we show the ratio of NLO (NNLO) leading over NLO $bg$ by blue (red) lines. }
    \label{fig:PIM_mass}
\end{figure}

\begin{figure}[t]
    \centering
    \includegraphics[width=0.49\textwidth]{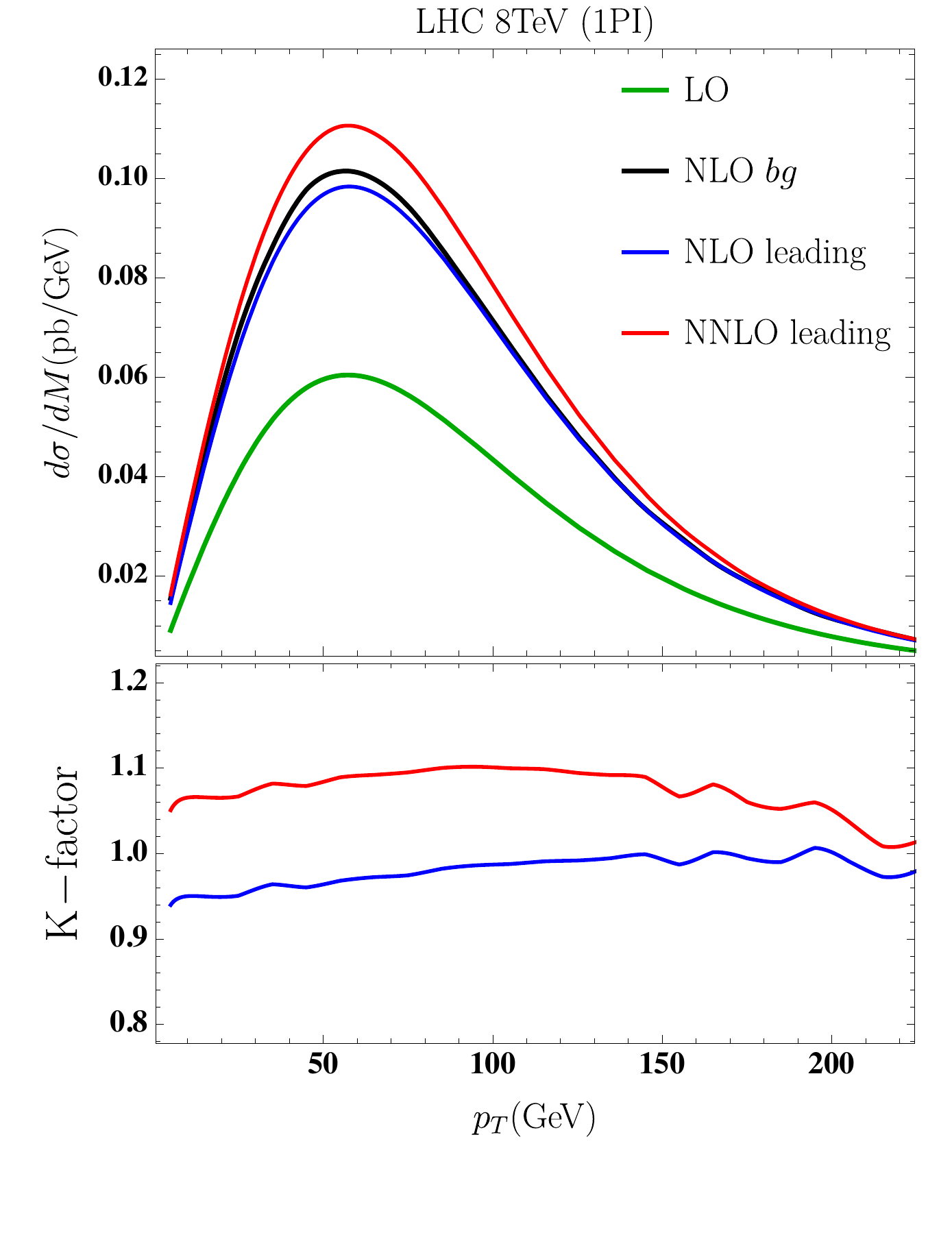}~
    \includegraphics[width=0.49\textwidth]{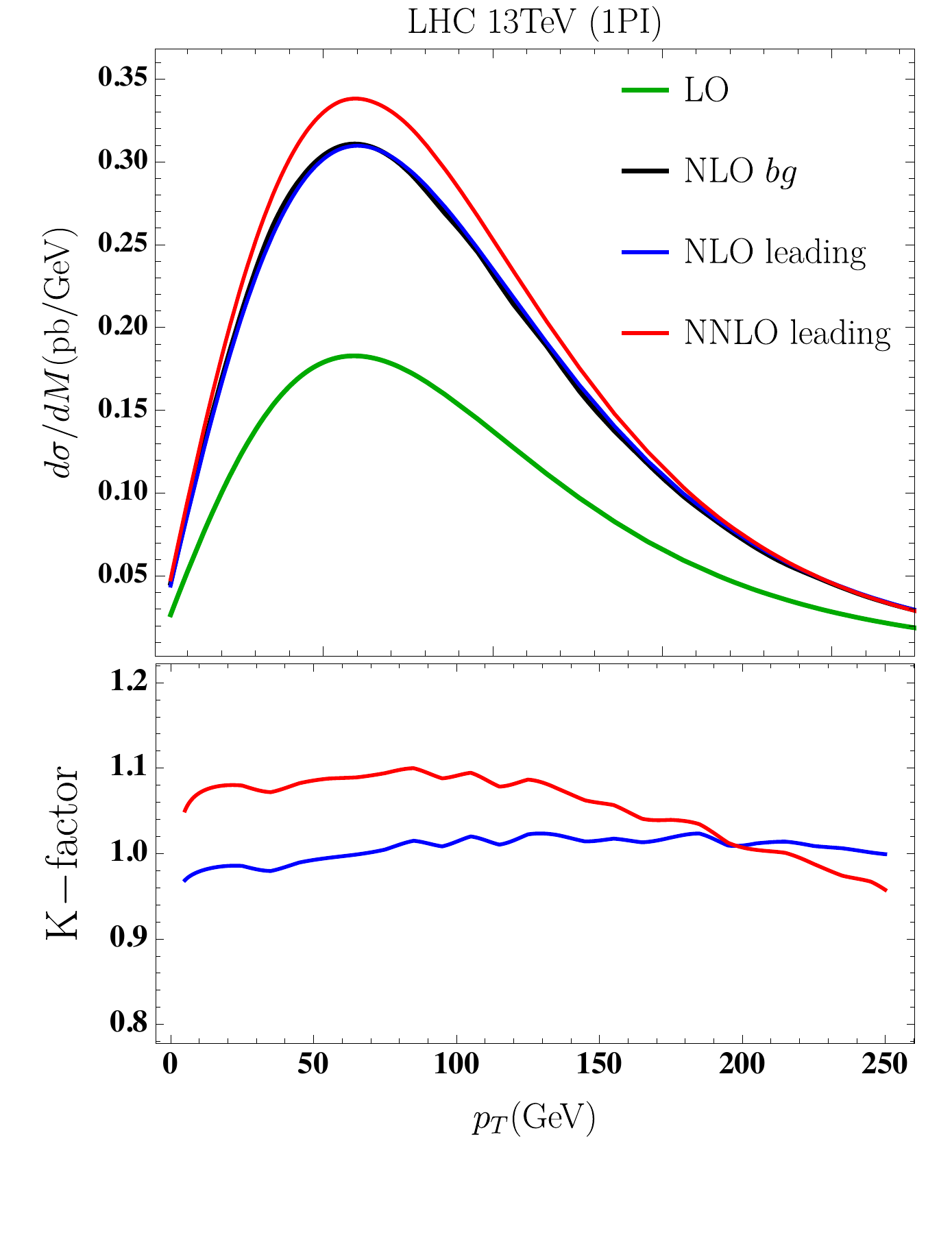}
    \vspace{-1.5cm}    
    \caption{Top quark $p_T$ distributions in the 1PI scheme  for $tW^-$ production. The color scheme is the same as figure~\ref{fig:PIM_mass}.
    }
    \label{fig:1PI_pt}
\end{figure}

Then we turn to the differential cross sections. 
We show the $tW$ invariant mass distributions in the PIM scheme in figure \ref{fig:PIM_mass} 
and the top quark $p_T$ distributions in the 1PI scheme in figure~\ref{fig:1PI_pt}.
We show results at both the 8 TeV and 13 TeV LHC.
It can be seen that the leading power terms are dominant in all the invariant mass or the top $p_T$ regions, as in the case of total cross sections.  
The NNLO leading terms increase the NLO leading cross section by about $10\%$ in  most of the region. 

\begin{table}[b]
    \centering
    \begin{tabular}{l|c|c|c|c}
    \hline\hline
     [pb]   & \multicolumn{2}{c|}{PIM  }   & \multicolumn{2}{c}{1PI}  \\
     \hline 
    $\sqrt{s}$ & 8 TeV & 13 TeV     &   8 TeV & 13 TeV  \\
     \hline
%     NLO $bg$ &  $12.0^{+2\%}_{-6\%}$ &   $39.4^{+4\%}_{-7\%}$   & $11.7^{+4\%}_{-8\%}$ & $38.2^{+6\%}_{-10\%}$
%      \\
%           \hline
%     NLO leading   & $11.2^{+2\%}_{-5\%}$ & $36.8^{+3\%}_{-6\%}$ &  $11.5^{+0\%}_{-3\%}$ &  $38.0^{+2\%}_{-5\%}$ 
%     \\
%    \hline
     LO  &  $7.0^{+5\%}_{-6\%}$ & $22.4^{+5\%}_{-2\%}$ & $7.2^{+5\%}_{-4\%}$  & $22.9^{+3\%}_{-1\%}$
      \\
            \hline
     NLO  &  $9.92^{+2\%}_{-2\%}$ & $32.8^{+1\%}_{-1\%}$ & $10.0^{+1\%}_{-2\%}$  & $33.0^{+1\%}_{-1\%}$
      \\
            \hline            
     aNNLO  & $11.6^{+4\%}_{-5\%}$  & $37.1^{+5\%}_{-5\%}$  &  $11.2^{+6\%}_{-6\%}$  & $35.9^{+7\%}_{-6\%}$ 
      \\
            \hline            
     NLO+NNLL & $11.4^{+7\%}_{-7\%}$& $36.7^{+7\%}_{-7\%}$ &$11.7^{+12\%}_{-17\%}$ & $37.3^{+16\%}_{-21\%}$
      \\
                  \hline            
     aNNLO/NLO & 1.16 & 1.13 & 1.12 & 1.09 
     \\
                  \hline            
     (NLO+NNLL)/NLO & 1.15 & 1.12 & 1.17 & 1.13 
      \\
     \hline \hline
    \end{tabular}
    \caption{Total cross sections for $tW^-$ production in PIM and 1PI schemes.  The NLO cross sections  are calculated using DS scheme. }
    \label{tab:sigma}
\end{table}

Now we present in table \ref{tab:sigma} the aNNLO and NLO+NNLL result defined in eq.~(\ref{eq:annlo}) and eq.~(\ref{eq:nnllnlo}), respectively. 
The NLO+NNLL (aNNLO) predictions increase the NLO total cross section by  $12\%\sim 17\%$ ($9\%\sim 16\%$) depending on the collider energy and the threshold variable, but with larger scale uncertainties. 
These large uncertainties are mainly from the variation of the factorization scale $\mu_f$. At first sight, this is unexpected since we have checked the scale independence near the threshold analytically in eq.~(\ref{eq:sclindep}). 
However, this is based on the assumption $x_{1,2}\to1$ as discussed in appendix~\ref{app:pdfs}.
When the kinematics is far away from the threshold limit, this assumption is not valid.
The very small scale uncertainties of the NLO results seem like a coincidence because the NLO contributions from $gg$ and $qq^\prime$ channels are negative while the contributions from $bg$ channel are positive. Meanwhile they display an opposite behavior under the  scale variation; see table \ref{tab:nlosigma}.
Our resummed result or its expansion in $\alpha_s$ improves only the result in $bg$ channel.
It would be interesting to investigate whether the scale cancellation among different channels happens  at higher orders.
From table \ref{tab:sigma} we also find that the total cross sections in the PIM and 1PI scheme are compatible.  
And the resummed cross sections in PIM kinematics have smaller scale uncertainties. 

\begin{figure}[t]
    \centering
    \hspace{-0.5cm}\includegraphics[width=0.7\textwidth]{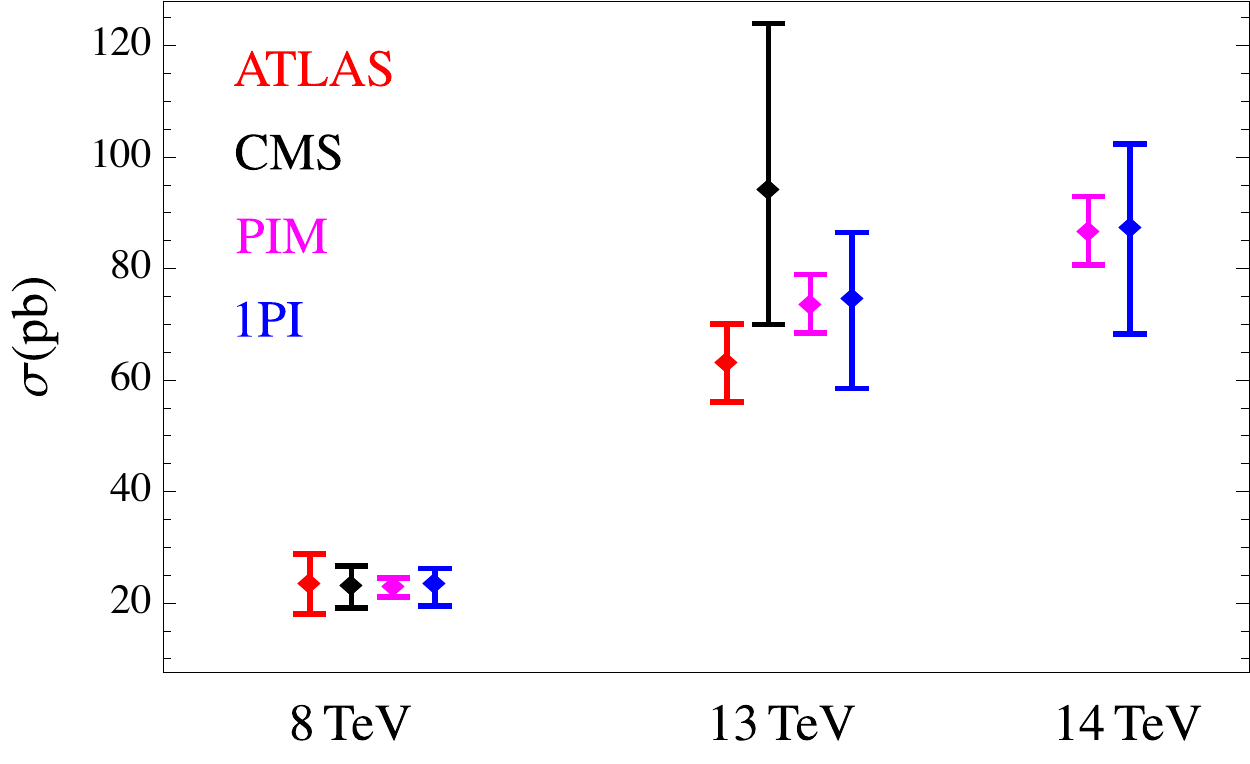}
    \caption{Comparison between measured cross section for $tW^-$ and $\bar{t}W^+$ production  at the LHC~\cite{Chatrchyan:2014tua,Aad:2015eto,Aaboud:2016lpj,Sirunyan:2018lcp} and RG-improved predictions.  }
    \label{fig:measu}
\end{figure}

Lastly, we compare the theoretical results with the measurements of the total cross section for $tW^-$ and $\bar tW^+$ production at the LHC in figure~\ref{fig:measu}. After considering the large experimental uncertainties, the NLO+NNLL predictions are in good  agreement with the data at the 8 TeV and 13 TeV LHC.  
We also give the predictions at the 14 TeV LHC.

\section{Conclusion} 
\label{conclu}

We have investigated the soft-gluon resummation for $tW$ production in the framework of soft-collinear effective theory.  
We considered the two different definitions of the threshold limit, $1-M^2/\hat{s} \to 0 $ and $ s_4 \to 0$, corresponding to the PIM and 1PI kinematic schemes, respectively.  
We briefly discussed the factorization and resummation formalism in both kinematic schemes.
In addition, we have calculated the hard function and  soft function at NLO. 
Expanding the resummed formula in $\alpha_s$ gives the leading power terms of the fixed-order results.
We found that the NLO leading power contribution is a good approximation to the  $bg$ channel at NLO not only for the total cross sections but also for the differential distributions.
After resumming the soft gluon effects to all orders using renormalization group equation, 
we find that the NLO+NNLL results
increase the NLO cross sections by about 15(12)\% in PIM and 17(13)\% in 1PI scheme at the 8(13) TeV LHC, but with large uncertainties which is mostly generated by varying the factorization scale. 
We compared with the data at the 8 and 13 TeV LHC and found good agreement within uncertainties. 
We provide the prediction for the 14 TeV LHC.

In future, we can obtain more precise predictions for the $tW$ process by including higher order hard and soft functions in the resummation formalism or by calculating the full NNLO corrections. 
The latter may be achieved making use of the $N$-jettiness subtraction method~\cite{Gao:2012ja,Gaunt:2015pea,Boughezal:2015dva}. 
The NNLO beam function~\cite{Gaunt:2014xga,Gaunt:2014cfa} and $N$-jettiness soft function~\cite{Li:2018tsq} for this process have been computed. 
The only missing part is the two-loop hard function, which requires  a huge amount of work. We defer this study to future work.

\section*{Acknowledgements}
We would like to thank Shi Ang Li for the contribution in the early stage of this work.  C.S.L. was supported by the National Nature  Science foundation of China, under Grants No. 11875072.  H.T.L. was supported by the Los Alamos National Laboratory LDRD program. J.W. was  supported  by the BMBF project No.  05H15WOCAA and 05H18WOCA1.

\appendix

\section{RG equation of the PDFs near the threshold} \label{app:pdfs}

In the threshold limit  $x_{1,2}\to 1$, the DGLAP evolution  for the PDFs can be written as \cite{Korchemsky:1992xv,Moch:2004pa}
\begin{align}
    \frac{d}{d\ln\mu} f_{i/p} (x,\mu) = \int_y^1 \frac{dz}{z} \left[ \frac{2 C_i \gamma^{\rm cusp}(\alpha_s)}{(1-z)_+} + 2 \gamma^{\phi_i}(\alpha_s) \delta(1-z) \right] f_{i/p} (x/z, \mu ), 
\end{align}
where the quadratic Casimir operator $C_i$ for the quark  is $C_q = C_F$, and for the gluon is
$C_g = C_A$.  In the threshold limit $s_4\to 0 $ the evolution equations for PDFs are
\begin{align}
    \frac{d}{d\ln\mu} f_{q/p}(x_1(s_4),\mu) & = 2 C_F \gamma^{\rm cusp}(\alpha_s)  \int_0^{s_4} ds_4^\prime \frac{ f_{q/p}(x_1(s^\prime_4),\mu)-f_{q/p}(x_1(s_4),\mu)}{s_4-s_4^\prime} 
    \nonumber \\
    &~~~ + \left[2 C_F \gamma^{\rm cusp}(\alpha_s)  \ln \frac{s_4}{-\hat{u}_1^W} + 2 \gamma^{\phi_q}(\alpha_s)\right]  f_{q/p}(x_1(s_4),\mu), 
     \nonumber \\ 
    \frac{d}{d\ln\mu} f_{g/p}(x_2(s_4),\mu) & = 2 C_A \gamma^{\rm cusp}(\alpha_s) \int_0^{s_4} ds_4^\prime  \frac{ f_{g/p}(x_2(s^\prime_4),\mu)-f_{g/p}(x_2(s_4),\mu)}{s_4-s_4^\prime} 
    \nonumber \\
    & ~~~+ \left[2 C_A \gamma^{\rm cusp}(\alpha_s)  \ln \frac{s_4}{-\hat{t}_1^{\,W}} + 2 \gamma^{\phi_g}(\alpha_s)\right]  f_{g/p}(x_2(s_4),\mu).
\end{align}
A similar derivation for $t\bar{t}$ and single top production can be found in refs.~\cite{Ahrens:2011mw, Wang:2012dc}.

\section{Soft function} \label{app:soft}
The NLO soft function can be written as 
\begin{align}
    \tilde{s}_{\rm NLO}(L,\mu_s) = \frac{\alpha_s}{4\pi}\bigg[-C_A I_{12} - (2\,C_F-C_A)I_{13} - C_A I_{23} + C_F I_{33}\bigg],
\end{align}
For convenience we evaluate soft integral $I_{ij}$ in the position space, and then transform them into Laplace space as
\begin{align}
    I_{ij}(L) = -\frac{(4\pi \mu)^{2\epsilon}}{\pi^{2-\epsilon}} v_i \cdot v_j \int d^d k \frac{e^{-i k^0 x_0}}{v_i \cdot k v_j \cdot k} (2\pi) \delta(k^2) \theta(k^0) \Bigg |_{L_0 \to - L},
\end{align}
with $L_0 =\ln\left(-\mu^2 x_0^2 e^{2\gamma_{\rm E}}/4 \right)$, and $v_i$ are normalized momenta fulfilling on-shell conditions as $v_1^2=v_2^2=0$ and $v_3^2=1$. 

In the PIM kinematics the full set of integrals can be found in section III of ref.~\cite{Li:2014ula}. In the 1PI kinematics the integral $I_{12}$ can be obtained from eq.~(21) of ref.~\cite{Ahrens:2011mw} by replacing all the kinematics variables to the ones related to $W$. The integrals $I_{13}$, $I_{23}$ and $I_{33}$ are more complicated, and have been first calculated in this paper.  The non-vanishing integrals are collected below 
\begin{align}
    I_{12} &= -\left( L + \ln \frac{\hat{s} M_W^2}{\hat{t}_1^W \hat{u}_1^W} \right)^2 - \frac{\pi^2}{6} - 2 \, {\rm Li}_2\left( 1-\frac{\hat{s} M_W^2}{ \hat{t}_1^W \hat{u}_1^W} \right),  \nonumber \\
    I_{13} &= - \frac{1}{2} \left(L + 2 \ln \frac{M_W \hat{t}_1}{m_t \hat{u}_1^W} \right)^2  - \frac{\pi^2}{12} - 2\, {\rm Li}_2\left(1-\frac{M_W}{m_t\, x_{ tW}}\frac{ \hat{t}_1}{ \hat{u}_1^W } \right)  
    - 2\, {\rm Li}_2\left(1-\frac{M_W \, x_{ tW}}{m_t} \frac{  \hat{t}_1}{ \hat{u}_1^W } \right),
    \nonumber \\
    I_{33} &= - 2 L - 4 \frac{1+\beta_t  \beta_W}{\beta_t+\beta_W} \ln x_{tW}, 
    \nonumber \\
    I_{23} &=  I_{13}(\hat{t}_1 \to \hat{u}_1, \hat{u}_1^W\to \hat{t}_1^W), 
\end{align}
with $\beta_t=\sqrt{1-4m_t^2\hat s /(m_t^2-M_W^2+\hat s)^2} $, $\beta_W =\sqrt{1-4M_W^2\hat s/(m_t^2 - M_W^2 - \hat s)^2}$, $x_{tW} = \sqrt{x_t\  x_W}$ and $x_i = (1-\beta_i)/(1+\beta_i)$. In the limit of $M_W\to m_t$, the integrals reproduce those for $t\bar{t}$ production in ref.~\cite{Ahrens:2011mw}.

\bibliographystyle{JHEP}

\bibliography{references}

\end{document}